\documentclass[a4paper,aps,twocolumn,nofootinbib]{revtex4}
\RequirePackage[english]{babel}
\RequirePackage[latin1]{inputenc}
\RequirePackage[T1]{fontenc}
\RequirePackage{mathrsfs}
\RequirePackage{amsmath}
\RequirePackage{amssymb}
\RequirePackage{amsbsy}
\RequirePackage{color}
\RequirePackage{bm}
\begin{document}
\title{\bf{Spinor Fields, Singular Structures, Charge Conjugation, ELKO\\
and Neutrino Masses}}
\author{Luca Fabbri}
\affiliation{DIME, Universit\`{a} di Genova, P.Kennedy Pad.D, 16129 Genova, ITALY}
\date{\today}
\begin{abstract}
In this paper, we consider the most general treatment of spinor fields, their kinematic classification and the ensuing dynamic polar reduction, for both classes of regular and singular spinors; specifying onto the singular class, we discuss features of the corresponding field equations, taking into special account the sub-classes of Weyl and Majorana spinors; for the latter case, we study the condition of charge-conjugation, presenting a detailed introduction to a newly-defined type of spinor, that is the so-called ELKO spinor: at the end of our investigation, we will assess how all elements will concur to lay the bases for a simple proposal of neutrino mass generation.
\end{abstract}
\maketitle
\section{Introduction}
In recent years, physics has witnessed the rise of a new dichotomy, this time internal to theoretical physics, and namely the separation between proper theoretical physics and the more specialized sector of mathematical physics: the former is still about the construction of models, like the standard model, while the latter is mainly about the mathematical study of the general properties of fields, as the examination of Clifford structures and the subsequent opportunity to address the definition of extended types of spinor fields and spinor field dynamics. This last case, the approach of mathematical physics, is motivated by a rationale of total generality, with the hope that in doing so we could unravel the existence of interesting and useful concepts, which might well remain hidden otherwise.

One of the most general analyses is carried on about spinors fields, due in part to their central role and in part to the fact that they are defined in terms of transformation properties, which make it possible to perform several forms of classifications: among these, the first and most known is due to Wigner, who used the momentum vector and the Pauli-Lubanski spin axial-vector to classify the spinor in terms of its mass and spin; also known is the Lounesto classification \cite{L}, which is performed employing the bi-linear invariants that can be constructed in terms of pair of adjoint spinors; other classifications have been done in more detailed fashion in \cite{Cavalcanti:2014wia} and \cite{Fabbri:2016msm}. All of these classifications are somehow complementary, although for our purposes we only consider the classification of \cite{Fabbri:2016msm}.

This classification employs only the definition of spinor given in terms of its (active) spinorial transformation law in the case in which its parameters are point-dependent (local) \cite{Rodrigues:2005yz}: a point-dependent spinorial transformation is the only one capable of having the point-dependent components of a spinor field transferred into point-dependent tetrad frames \cite{Fabbri:2016msm}. Therefore, it is possible to have spinor fields stripped of even more of their components, if we are willing to pay the price of a non-null spinorial connection in the spinorial covariant derivatives. This might sound a strange mathematical trick, but it is just the spinorial equivalent of examining a spinning top by going into the frame that rotates with it. In such frame the study of the top would clearly be simplified, although now one would have to account also for centrifugal inertial effects.

According to this analysis then, it becomes possible to see that spinors can be split into two classes according to whether their spinorial bi-linear scalar and pseudo-scalar are both equal to zero or not: the latter are the regular spinors and they have been studied in \cite{Fabbri:2017pwp}; the former are the singular spinors and they are the subject of this work.

It is important to notice that inside each class, there can be further sub-classes: for instance, in the class that contains singular spinors, we find both Weyl spinors and Majorana spinors \cite{Fabbri:2016msm}. This last case will be the subject of a deeper investigation about the general conditions of charge-conjugation, especially regarding analogies with a type of recently-introduced spinor called ELKO \cite{Ahluwalia:2004sz,Ahluwalia:2004ab}.

As a result of the comparison between ELKO and Majorana spinors, we will conclude with some comment on a solution to the problem of neutrino mass generation.
\section{Geometrical Spinor Fields}
To begin, we recall the basic theory of spinor fields \cite{Fabbri:2016msm}.

The metric tensor is given by $g_{\alpha\rho}$ and it will be used to move coordinate indices; tetrads $e^{\alpha}_{a}$ are always taken to be ortho-normal $g_{\alpha\rho}e^{\alpha}_{a}e^{\rho}_{b}\!=\!\eta_{ab}$ and used to pass from coordinate (Greek) indices to Lorentz (Latin) indices, thus the Minkowskian matrix $\eta_{ab}$ is used to move the Lorentz indices; matrices $\boldsymbol{\gamma}^{a}$ belong to the Clifford algebra, so we may define $\left[\boldsymbol{\gamma}^{a}\!,\!\boldsymbol{\gamma}^{b}\right]\!=\! 4\boldsymbol{\sigma}^{ab}$ and $2i\boldsymbol{\sigma}_{ab}\!=\!\varepsilon_{abcd}\boldsymbol{\pi}\boldsymbol{\sigma}^{cd}$ where the parity-odd matrix $\boldsymbol{\pi}$ is implicitly defined\footnote{This is usually indicated as gamma with index five, but since this index has no meaning, we use a notation with no index.}, and $\boldsymbol{\gamma}_{0}$ is used to define the conjugate spinor $\overline{\psi}\!=\!\psi^{\dagger}\boldsymbol{\gamma}_{0}$ so that
\begin{eqnarray}
&2i\overline{\psi}\boldsymbol{\sigma}^{ab}\psi\!=\!M^{ab}\\
&\overline{\psi}\boldsymbol{\gamma}^{a}\boldsymbol{\pi}\psi\!=\!S^{a}\\
&\overline{\psi}\boldsymbol{\gamma}^{a}\psi\!=\!U^{a}\\
&i\overline{\psi}\boldsymbol{\pi}\psi\!=\!\Theta\\
&\overline{\psi}\psi\!=\!\Phi
\end{eqnarray}
are all real-valued bi-linear tensor quantities. They verify
\begin{eqnarray}
&M_{ab}\Phi\!+\!\frac{1}{2}\varepsilon_{abik}M^{ik}\Theta\!=\!U^{j}S^{k}\varepsilon_{jkab}\\
&M_{ab}\Theta\!-\!\frac{1}{2}\varepsilon_{abik}M^{ik}\Phi\!=\!U_{[a}S_{b]}
\end{eqnarray}
together with
\begin{eqnarray}
&M_{ik}U^{i}=\Theta S_{k}\label{P1}\\
&-\frac{1}{2}\varepsilon_{abik}M^{ab}U^{i}\!=\!\Phi S_{k}\label{L1}\\
&M_{ik}S^{i}=\Theta U_{k}\label{P2}\\
&-\frac{1}{2}\varepsilon_{abik}M^{ab}S^{i}\!=\!\Phi U_{k}\label{L2}
\end{eqnarray}
and
\begin{eqnarray}
&\frac{1}{2}M_{ab}M^{ab}\!=\!\Phi^{2}\!-\!\Theta^{2}\label{norm2}\\
&U_{a}U^{a}\!=\!-S_{a}S^{a}\!=\!\Theta^{2}\!+\!\Phi^{2}\label{norm1}\\
&\frac{1}{4}M_{ab}M_{ij}\varepsilon^{abij}\!=\!2\Theta\Phi\label{orthogonal2}\\
&S_{a}U^{a}\!=\!0\label{orthogonal1}
\end{eqnarray}
called Fierz re-arrangements, and being spinor identities.

To classify spinorial fields, we first consider the case in which at least one of the two scalars $\Theta$ or $\Phi$ is non-zero: in this situation, (\ref{norm1}) tells that $U^{a}$ is time-like, and so it is always possible to perform up to three boosts in order to bring its space components to vanish; then, it is always possible to use the rotations around the first and second axis in order to bring the space part of $S^{a}$ aligned with the third axis; finally it is always possible to employ the third rotation in order to bring the spinor into the form
\begin{eqnarray}
&\!\!\psi\!=\!\boldsymbol{S}\!\left(\!\begin{tabular}{c}
$e^{i\frac{\beta}{2}}$\\
$0$\\
$e^{-i\frac{\beta}{2}}$\\
$0$
\end{tabular}\!\right)\phi
\label{regular}
\end{eqnarray}
up to a transformation of type $\psi\!\rightarrow\!\boldsymbol{\pi}\psi$ and up to a third axis reflection, and where $\boldsymbol{S}$ is a general spinorial transformation law. The complementary situation is presented whenever both the scalars $\Theta$ and $\Phi$ are identically zero: in this case, (\ref{norm2}, \ref{orthogonal2}) tell that if $M_{ab}$ is written in terms of components $M_{0K}\!=\!E_{K}$ and $M_{IJ}\!=\!\varepsilon_{IJK}B^{K}$ then these components verify $E^{2}\!=\!B^{2}$ and $\vec{E}\!\cdot\!\vec{B}\!=\!0$ and this implies that it is always possible to perform up to three rotations in order to bring $\vec{B}$ and $\vec{E}$ aligned respectively with first and second axis; then, it is always possible to employ the third boost in order to bring the spinor into the form
\begin{eqnarray}
&\!\!\psi\!=\!\boldsymbol{S}e^{i\chi}\!\left(\!\begin{tabular}{c}
$\cos{\frac{\theta}{2}}$\\
$0$\\
$0$\\
$\sin{\frac{\theta}{2}}$
\end{tabular}\!\right)
\label{singular}
\end{eqnarray}
up to a transformation of type $\psi\!\rightarrow\!\boldsymbol{\pi}\psi$ and up to a third axis reflection, with $\boldsymbol{S}$ being a general spinorial transformation law. Because of the complementarity of these two classes, any spinor must be in either of them \cite{Fabbri:2016msm}.

From the metric, we define $\Lambda^{\sigma}_{\alpha\nu}$ as the symmetric connection; with it, the tetrad fields and the Minkowski matrix, we define $\Omega^{ak}_{\phantom{ak}\pi}\!=\!\eta^{kb}
\xi^{\nu}_{b}\xi^{a}_{\sigma} (\Lambda^{\sigma}_{\nu\pi}\!\!-\!\xi^{\sigma}_{i}\partial_{\pi}\xi_{\nu}^{i})\!=\!-\Omega^{ka}_{\phantom{ka}\pi}$ as the spin connection; with it, and together with the gauge potentials $A_{\mu}$ with $q$ charge, it is finally possible to define
\begin{eqnarray}
&\boldsymbol{\Omega}_{\mu}
=\frac{1}{2}\Omega^{ab}_{\phantom{ab}\mu}\boldsymbol{\sigma}_{ab}
\!+\!iqA_{\mu}\boldsymbol{\mathbb{I}}\label{spinorialconnection}
\end{eqnarray}
called spinorial connection and in terms of which we can define $\boldsymbol{\nabla}_{\mu}\psi\!=\! \partial_{\mu}\psi\!+\!\boldsymbol{\Omega}_{\mu}\psi$ to be the torsionless form of the spinorial covariant derivative. In giving this definition we assumed the coordinate connection to be symmetric, that is torsionless, but this amounts to no loss of generality, if torsion is eventually added in the form of the completely antisymmetric dual of an axial-vector field $W_{\sigma}$ \cite{Fabbri:2014dxa}.

According to the above classification, (\ref{regular}) would give rise to the spinorial covariant derivative of the form
\begin{eqnarray}
\nonumber
&\boldsymbol{\nabla}_{\mu}\psi\!=\![\nabla_{\mu}\ln{\phi}\mathbb{I}-\\
\nonumber
&-\frac{i}{2}\nabla_{\mu}\beta\boldsymbol{\pi}+\\
\nonumber
&+i(qA_{\mu}\!-\!P_{\mu})\mathbb{I}+\\
&+\frac{1}{2}(\Omega_{ij\mu}\!-\!R_{ij\mu})\boldsymbol{\sigma}^{ij}]\psi
\label{decspinderreg}
\end{eqnarray}
in which $\boldsymbol{S}\partial_{\mu}\boldsymbol{S}^{-1}\!=\!iP_{\mu}\mathbb{I}
+\frac{1}{2}R_{ij\mu}\boldsymbol{\sigma}^{ij}$ with $P_{\mu}$ gauge vector and $R_{\alpha\nu\mu}$ spin connection, and with $A_{\mu}$ and $\Omega_{ij\mu}$ as the electrodynamic potential and gravitational field; remark that because the spinor field is point-dependent, having components transferred into the frame requires that the transformation law $\boldsymbol{S}$ be also point-dependent, and this means in turn that the spinorial connection has non-zero components: the balance between degrees of freedom and non-zero components of the spinorial connection is shown in the fact that the spinorial covariant derivative is indeed covariant term by term. In the complementary situation, we would have that (\ref{singular}) would give rise to the covariant spinorial derivative according to the following form
\begin{eqnarray}
\nonumber
&\boldsymbol{\nabla}_{\mu}\psi\!=\![-\frac{i}{2}\nabla_{\mu}\theta 
\boldsymbol{S}\boldsymbol{\gamma}^{2}\boldsymbol{S}^{-1}\boldsymbol{\pi}+\\
\nonumber
&+i\nabla_{\mu}\chi+\\
\nonumber
&+i(qA_{\mu}\!-\!P_{\mu})\mathbb{I}+\\
&+\frac{1}{2}(\Omega_{ij\mu}\!-\!R_{ij\mu})\boldsymbol{\sigma}^{ij}]\psi
\label{decspindersing}
\end{eqnarray}
in which a strange term $\boldsymbol{S}\boldsymbol{\gamma}^{2}\boldsymbol{S}^{-1}$ has appeared: it is quite difficult to interpret this term but one thing we may say is that it clearly depends on the frame. These forms are called polar forms of the spinorial covariant derivative.

The dynamics is set by the complete Lagrangian
\begin{eqnarray}
\nonumber
&\mathscr{L}\!=\!\frac{1}{4}(\partial W)^{2}\!-\!\frac{1}{2}M^{2}W^{2}
\!+\!R\!+\!\frac{1}{4}F^{2}-\\
&-i\overline{\psi}\boldsymbol{\gamma}^{\mu}\boldsymbol{\nabla}_{\mu}\psi
\!+\!X\overline{\psi}\boldsymbol{\gamma}^{\mu}\boldsymbol{\pi}\psi W_{\mu}
\!+\!m\overline{\psi}\psi
\label{l}
\end{eqnarray}
in which $\partial W$ is the curl of the axial-vector torsion, $R$ the Ricci scalar and $F$ the Maxwell tensor, and where $X$, $M$ and $m$ are the torsion-spinor coupling constant, and the torsion and spinor masses, considered for generality. 

However, this full Lagrangian is not really going to be needed in the following, because we intend to specialize on the Dirac spinor field, whose spinor field equation is
\begin{eqnarray}
&i\boldsymbol{\gamma}^{\mu}\boldsymbol{\nabla}_{\mu}\psi
\!-\!XW_{\sigma}\boldsymbol{\gamma}^{\sigma}\boldsymbol{\pi}\psi\!-\!m\psi\!=\!0
\label{D}
\end{eqnarray}
which is the equation we work out next. Multiplying this equation by the Clifford matrices and the adjoint spinor and then splitting real and imaginary parts, we obtain
\begin{eqnarray}
&\frac{i}{2}(\overline{\psi}\boldsymbol{\gamma}^{\mu}\boldsymbol{\nabla}_{\mu}\psi
\!-\!\boldsymbol{\nabla}_{\mu}\overline{\psi}\boldsymbol{\gamma}^{\mu}\psi)
\!-\!XW_{\sigma}S^{\sigma}\!-\!m\Phi\!=\!0\\
&\nabla_{\mu}U^{\mu}\!=\!0
\end{eqnarray}
\begin{eqnarray}
&\frac{i}{2}(\overline{\psi}\boldsymbol{\gamma}^{\mu}\boldsymbol{\pi}\boldsymbol{\nabla}_{\mu}\psi
\!-\!\boldsymbol{\nabla}_{\mu}\overline{\psi}\boldsymbol{\gamma}^{\mu}\boldsymbol{\pi}\psi)
\!-\!XW_{\sigma}U^{\sigma}\!=\!0\\
&\nabla_{\mu}S^{\mu}\!-\!2m\Theta\!=\!0
\end{eqnarray}
\begin{eqnarray}
\nonumber
&\frac{i}{2}(\overline{\psi}\boldsymbol{\nabla}^{\alpha}\psi
\!-\!\boldsymbol{\nabla}^{\alpha}\overline{\psi}\psi)
\!-\!\frac{1}{2}\nabla_{\mu}M^{\mu\alpha}-\\
&-\frac{1}{2}XW_{\sigma}M_{\mu\nu}\varepsilon^{\mu\nu\sigma\alpha}\!-\!mU^{\alpha}\!=\!0
\label{vr}\\
\nonumber
&\nabla_{\alpha}\Phi
\!-\!2(\overline{\psi}\boldsymbol{\sigma}_{\mu\alpha}\!\boldsymbol{\nabla}^{\mu}\psi
\!-\!\!\boldsymbol{\nabla}^{\mu}\overline{\psi}\boldsymbol{\sigma}_{\mu\alpha}\psi)+\\
&+2X\Theta W_{\alpha}\!=\!0\label{vi}
\end{eqnarray}
\begin{eqnarray}
\nonumber
&\nabla_{\nu}\Theta\!-\!
2i(\overline{\psi}\boldsymbol{\sigma}_{\mu\nu}\boldsymbol{\pi}\boldsymbol{\nabla}^{\mu}\psi\!-\!
\boldsymbol{\nabla}^{\mu}\overline{\psi}\boldsymbol{\sigma}_{\mu\nu}\boldsymbol{\pi}\psi)-\\
&-2X\Phi W_{\nu}\!+\!2mS_{\nu}\!=\!0\label{ar}\\
\nonumber
&(\boldsymbol{\nabla}_{\alpha}\overline{\psi}\boldsymbol{\pi}\psi
\!-\!\overline{\psi}\boldsymbol{\pi}\boldsymbol{\nabla}_{\alpha}\psi)
\!-\!\frac{1}{2}\nabla^{\mu}M^{\rho\sigma}\varepsilon_{\rho\sigma\mu\alpha}+\\
&+2XW^{\mu}M_{\mu\alpha}\!=\!0\label{ai}
\end{eqnarray}
\begin{eqnarray}
\nonumber
&\nabla^{\mu}S^{\rho}\varepsilon_{\mu\rho\alpha\nu}
\!+\!i(\overline{\psi}\boldsymbol{\gamma}_{[\alpha}\!\boldsymbol{\nabla}_{\nu]}\psi
\!-\!\!\boldsymbol{\nabla}_{[\nu}\overline{\psi}\boldsymbol{\gamma}_{\alpha]}\psi)+\\
&+2XW_{[\alpha}S_{\nu]}\!=\!0\\
\nonumber
&\nabla^{[\alpha}U^{\nu]}\!+\!i\varepsilon^{\alpha\nu\mu\rho}
(\overline{\psi}\boldsymbol{\gamma}_{\rho}\boldsymbol{\pi}\!\boldsymbol{\nabla}_{\mu}\psi\!-\!\!
\boldsymbol{\nabla}_{\mu}\overline{\psi}\boldsymbol{\gamma}_{\rho}\boldsymbol{\pi}\psi)-\\
&-2XW_{\sigma}U_{\rho}\varepsilon^{\alpha\nu\sigma\rho}\!-\!2mM^{\alpha\nu}\!=\!0
\end{eqnarray}
which are known as Gordon-Madelung decompositions.

Having the above polar form of the spinorial covariant derivative (\ref{decspinderreg}) and all Gordon-Madelung decompositions it is possible to compute the corresponding polar form of the Gordon-Madelung decompositions; however, this procedure is highly redundant as all resulting expressions can be derived from the two following expressions
\begin{eqnarray}
\nonumber
&\frac{1}{2}\varepsilon_{\mu\alpha\nu\iota}(R\!-\!\Omega)^{\alpha\nu\iota}
\!-\!2(P\!-\!qA)^{\iota}u_{[\iota}s_{\mu]}-\\
&-2XW_{\mu}\!+\!\nabla_{\mu}\beta\!+\!2s_{\mu}m\cos{\beta}\!=\!0\label{f1}\\
\nonumber
&(R\!-\!\Omega)_{\mu a}^{\phantom{\mu a}a}
\!-\!2(P\!-\!qA)^{\rho}u^{\nu}s^{\alpha}\varepsilon_{\mu\rho\nu\alpha}+\\
&+2s_{\mu}m\sin{\beta}\!+\!\nabla_{\mu}\ln{\phi^{2}}\!=\!0\label{f2}
\end{eqnarray}
as the only two independent equations. As for the polar form of the spinorial covariant derivative (\ref{decspindersing}) it would in principle be possible to plug it in all Gordon-Madelung decompositions, but before embarking on such laborious task it would be desirable to understand the significance of the strange frame-related term. We shall postpone this discussion to the next section, entirely devoted to it.

For the moment, we would like to comment on the fact that what we called regular spinors are described within the Lounesto classification as the first three classes: the first class contains the most general spinor, and perhaps paradoxically this is the only class of which we know no exact solution in presence of interactions; the second class is characterized by the constraint $\Theta\!=\!0$ and in this class there are exact solutions in presence of gravitational field as described in \cite{Vignolo:2011qt} as well as standard solutions in plane waves for free particles in QFT; the third class is the one in which $\Phi\!=\!0$ and as weird as this may look nevertheless we do know exact solutions with gravity \cite{Cianci:2016pvd}. And what we called singular spinors are described by the Lounesto classification as the last three classes. We will undertake the discussion of these three classes in what follows.
\section{Classes of Singular Structure}
From now on we will focus on the classes containing singular spinors, those subject to the condition for which both scalars have $\Theta\!=\!\Phi\!=\!0$ identically: consequently, the remaining bi-linear spinors are such that the vector and axial vector are parallel and light-like $U_{a}U^{a}\!=\!0$ while for the tensor $M_{ab}M^{ab}\!=\!0$ and $M_{ab}M_{ij}\varepsilon^{abij}\!=\!0$ as well, and we also have $M_{ik}U^{i}=0$ and $\varepsilon_{abik}M^{ab}U^{i}\!=\!0$ as the Fierz conditions; as shown above, an explicit representation is given with $S_{k}\!=\!-\cos{\theta}U_{k}$ according to
\begin{eqnarray}
&U^{0}\!=\!1\\
&U^{3}\!=\!-1\\
&M^{02}\!=\!\sin{\theta}\\
&M^{23}\!=\!\sin{\theta}
\end{eqnarray}
obtained when
\begin{eqnarray}
&\!\!\psi\!=\!e^{i\chi}\!\left(\!\begin{tabular}{c}
$\cos{\frac{\theta}{2}}$\\
$0$\\
$0$\\
$\sin{\frac{\theta}{2}}$
\end{tabular}\!\right)
\end{eqnarray}
up to the transformation $\psi\!\rightarrow\!\boldsymbol{\pi}\psi$ and up to a third axis reflection, and valid in a specific frame. In the Lounesto classification these are further split into three classes: the fourth class contains spinors that have no supplementary constraint; the fifth class contains spinors subject to the further restriction $S^{a}\!=\!0$ such as Majorana spinors; the sixth class contains spinors subject to the further restriction $M^{ab}\!=\!0$ such as Weyl spinors. We notice that there can be no class for which $U^{a}\!=\!0$ because in this case the spinor field would simply be identically equal to zero.

Now, as we already had the opportunity to remark, in the expression of the spinorial covariant derivative
\begin{eqnarray}
\nonumber
&\boldsymbol{\nabla}_{\mu}\psi\!=\![-\frac{i}{2}\nabla_{\mu}\theta 
\boldsymbol{S}\boldsymbol{\gamma}^{2}\boldsymbol{S}^{-1}\boldsymbol{\pi}+\\
\nonumber
&+i\nabla_{\mu}\chi+\\
\nonumber
&+i(qA_{\mu}\!-\!P_{\mu})\mathbb{I}+\\
&+\frac{1}{2}(\Omega_{ij\mu}\!-\!R_{ij\mu})\boldsymbol{\sigma}^{ij}]\psi
\end{eqnarray}
we had the strange term $\boldsymbol{S}\boldsymbol{\gamma}^{2}\boldsymbol{S}^{-1}$ which was suggesting that this expression was also frame dependent. However, there is something even odder in this expression when we consider the Lounesto classification: since the $\theta$ angle is supposed to be a dynamical variable, then in principle it should be free to move across its values, and in particular passing from $\theta\!=\!0,\pi$ to $\theta\!=\!\pm\pi/2$ at will; because the first two values $\theta\!=\!0,\pi$ are those identifying Weyl spinors and the second two values $\theta\!=\!\pm\pi/2$ are those that identify a Majorana spinor, then it follows that it could be possible to have Weyl spinors mutating into a Majorana spinor or viceversa. This situation may look intriguing, although admittedly it does not appear to have a physical counterpart; another option may be that some mechanism would intervene to seal off two different Lounesto classes, hence removing the possibility that spinors of one class could mutate into spinors of another class. The last occurrence, if ever justified, would be implemented by the constraint that the angle should be constant, and a consequence of this fact would be that we would no longer have to worry about the strange frame-dependent factor. We also have to notice that for Weyl spinors we can always employ a combination of rotations to remove the phase; and for a Majorana spinor there can be no electrodynamic interaction and thus no phase is even admitted. Consequently, we may also neglect the phase in both cases. All elements put together, we have that it is a reasonable working hypothesis to assume the spinorial covariant derivative
\begin{eqnarray}
\nonumber
&\boldsymbol{\nabla}_{\mu}\psi\!=\![i(qA_{\mu}\!-\!P_{\mu})\mathbb{I}+\\
&+\frac{1}{2}(\Omega_{ij\mu}\!-\!R_{ij\mu})\boldsymbol{\sigma}^{ij}]\psi
\label{aux}
\end{eqnarray}
as the form we will employ in all of the following sections.

All Gordon-Madelung decompositions with such polar form (\ref{aux}) are consequently reduced to the expressions
\begin{eqnarray}
&[\frac{1}{4}(\Omega\!-\!R)^{\alpha\nu\rho}\varepsilon_{\alpha\nu\rho\mu}
\!-\!XW_{\mu}]S^{\mu}\!-\!(qA\!-\!P)_{\mu}U^{\mu}\!=\!0\\
&(\Omega\!-\!R)_{\mu a}^{\phantom{\mu a}a}U^{\mu}\!=\!0
\end{eqnarray}
\begin{eqnarray}
&[\frac{1}{4}(\Omega\!-\!R)^{\alpha\nu\rho}\varepsilon_{\alpha\nu\rho\mu}
\!-\!XW_{\mu}]U^{\mu}\!-\!(qA\!-\!P)_{\mu}S^{\mu}\!=\!0\\
&(\Omega\!-\!R)_{\mu a}^{\phantom{\mu a}a}S^{\mu}\!=\!0
\end{eqnarray}
\begin{eqnarray}
\nonumber
&\!\!\frac{1}{2}[\frac{1}{4}(\Omega\!-\!R)^{\alpha\nu\rho}\varepsilon_{\alpha\nu\rho\mu}\varepsilon^{\pi\kappa\mu\sigma}
\!-\!XW_{\mu}\varepsilon^{\pi\kappa\mu\sigma}+\\
&+g^{\sigma\kappa}(\Omega\!-\!R)^{\pi a}_{\phantom{\mu a}a}]M_{\pi\kappa}\!-\!mU^{\sigma}\!=\!0\\
&(qA\!-\!P)^{\mu}M_{\mu\sigma}\!=\!0
\end{eqnarray}
\begin{eqnarray}
&-\frac{1}{2}(qA\!-\!P)^{\mu}M^{\pi\kappa}\varepsilon_{\pi\kappa\mu\sigma}\!+\!mS_{\sigma}\!=\!0\\
\nonumber
&[\frac{1}{4}(\Omega\!-\!R)^{\alpha\nu\rho}\varepsilon_{\alpha\nu\rho\pi}g_{\kappa\sigma}
\!-\!XW_{\pi}g_{\kappa\sigma}-\\
&-\frac{1}{4}(\Omega\!-\!R)^{\mu a}_{\phantom{\mu a}a}\varepsilon_{\mu\pi\kappa\sigma}] M^{\pi\kappa}\!=\!0
\end{eqnarray}
\begin{eqnarray}
\nonumber
&[(\Omega\!-\!R)^{\rho\alpha\nu}\!+\!(\Omega\!-\!R)^{\nu\rho\alpha}
\!+\!(\Omega\!-\!R)^{\alpha\nu\rho}+\\
\nonumber
&+g^{\rho\alpha}(\Omega\!-\!R)^{\nu\mu}_{\phantom{\nu\mu}\mu}
\!-\!g^{\rho\nu}(\Omega\!-\!R)^{\alpha\mu}_{\phantom{\alpha\mu}\mu}-\\
&-2XW_{\sigma}\varepsilon^{\alpha\nu\sigma\rho}]S_{\rho}
\!-\!2(qA\!-\!P)_{\mu}U_{\rho}\varepsilon^{\alpha\nu\mu\rho}\!=\!0\\
\nonumber
&[(\Omega\!-\!R)^{\rho\alpha\nu}\!+\!(\Omega\!-\!R)^{\nu\rho\alpha}
\!+\!(\Omega\!-\!R)^{\alpha\nu\rho}+\\
\nonumber
&+g^{\rho\alpha}(\Omega\!-\!R)^{\nu\mu}_{\phantom{\nu\mu}\mu}
\!-\!g^{\rho\nu}(\Omega\!-\!R)^{\alpha\mu}_{\phantom{\alpha\mu}\mu}-\\
&\!\!\!\!-2X\!W_{\sigma}\varepsilon^{\alpha\nu\sigma\rho}]U_{\rho}
\!\!-\!2(qA\!-\!P)_{\mu}S_{\rho}\varepsilon^{\alpha\nu\mu\rho}\!\!-\!2m\!M^{\alpha\nu}\!\!=\!0
\end{eqnarray}
which in the following we discuss in terms of two disjoint sub-classes: one is given by Weyl, single-handed (either left-handed or right-handed), spinors, the other is given by Majorana, charged-conjugated (either self-conjugated or antiself-conjugated), spinors. We proceed in parallel.

The case of the Weyl spinor is given by single-handed, that is either left-handed or right-handed, spinorial fields: they have $M^{ab}\!=\!0$ so that $S^{a}\!=\!\pm U^{a}$ realized by
\begin{eqnarray}
&U_{0}\!=\!U_{3}\!=\!1
\end{eqnarray}
obtained when
\begin{eqnarray}
\psi\!=\!\left(\!\begin{tabular}{c}
$1$\\
$0$\\
$0$\\
$0$
\end{tabular}\!\right)\ \ \ \ \ \ \ \ \mathrm{or}\ \ \ \ \ \ \ \ 
\psi\!=\!\left(\!\begin{tabular}{c}
$0$\\
$0$\\
$1$\\
$0$
\end{tabular}\!\right)
\end{eqnarray}
for left-handed and right-handed cases, up to $\psi\!\rightarrow\!\boldsymbol{\pi}\psi$ and up to a third axis reflection, and valid only in a specific frame of reference. The complementary case of Majorana spinor is given when $\eta\boldsymbol{\gamma}^{2}\psi^{\ast}\!=\!\psi$ where $\eta$ is an irrelevant constant phase: in this case we have that $S^{a}\!=\!0$ and
\begin{eqnarray}
&U_{0}\!=\!U_{3}\!=\!1\ \ \ \ \ \ \ \ M^{02}\!=\!M^{23}\!=\!1
\end{eqnarray}
obtained when
\begin{eqnarray}
&\!\!\psi\!=\!\frac{1}{\sqrt{2}}\left(\!\begin{tabular}{c}
$1$\\
$0$\\
$0$\\
$1$
\end{tabular}\!\right)
\end{eqnarray}
up to the transformation $\psi\!\rightarrow\!\boldsymbol{\pi}\psi$ and up to a third axis reflection, and again valid in a specific frame. Both cases turn out to be spinors with a completely constant form.

The Gordon-Madelung decompositions in polar form in the Weyl case give that $m\!\equiv\!0$ identically and
\begin{eqnarray}
&(\Omega\!-\!R)_{\mu a}^{\phantom{\mu a}a}U^{\mu}\!=\!0\\
&[\frac{1}{4}(\Omega\!-\!R)^{\alpha\nu\rho}\varepsilon_{\alpha\nu\rho\mu}
\!-\!XW_{\mu}\!\pm\!(qA\!-\!P)_{\mu}]U^{\mu}\!=\!0\\
\nonumber
&[(\Omega\!-\!R)^{\rho\alpha\nu}\!+\!(\Omega\!-\!R)^{\nu\rho\alpha}
\!+\!(\Omega\!-\!R)^{\alpha\nu\rho}+\\
\nonumber
&+g^{\rho\alpha}(\Omega\!-\!R)^{\nu\mu}_{\phantom{\nu\mu}\mu}
\!-\!g^{\rho\nu}(\Omega\!-\!R)^{\alpha\mu}_{\phantom{\alpha\mu}\mu}-\\
&-2XW_{\sigma}\varepsilon^{\alpha\nu\sigma\rho}
\!\pm\!2(qA\!-\!P)_{\mu}\varepsilon^{\alpha\nu\mu\rho}]U_{\rho}\!=\!0
\end{eqnarray}
in terms of the vector current. For the Gordon-Madelung decompositions in polar form in the Majorana case there are considerable simplifications due to the fact that these spinors have no charge vector current nor spin axial vector current and so they do not couple to electrodynamics nor torsion, with the resulting equations of the form
\begin{eqnarray}
&(\Omega\!-\!R)_{\mu a}^{\phantom{\mu a}a}U^{\mu}\!=\!0\label{1}\\
&(\Omega\!-\!R)^{\alpha\nu\rho}U^{\mu}\varepsilon_{\alpha\nu\rho\mu}\!=\!0
\label{2}\\
\nonumber
&[(\Omega\!-\!R)^{\rho\alpha\nu}\!+\!(\Omega\!-\!R)^{\nu\rho\alpha}
\!+\!(\Omega\!-\!R)^{\alpha\nu\rho}+\\
&+g^{\rho\alpha}(\Omega\!-\!R)^{\nu\mu}_{\phantom{\nu\mu}\mu}
\!-\!g^{\rho\nu}(\Omega\!-\!R)^{\alpha\mu}_{\phantom{\alpha\mu}\mu}]U_{\rho}
\!\!-\!2m\!M^{\alpha\nu}\!\!=\!0\label{3}
\end{eqnarray}
similar to those of Weyl plus
\begin{eqnarray}
&[(\Omega\!-\!R)^{\alpha\nu\rho}\varepsilon_{\alpha\nu\rho\pi}g_{\kappa\sigma}\!-\!
(\Omega\!-\!R)^{\mu a}_{\phantom{\mu a}a}\varepsilon_{\mu\pi\kappa\sigma}]M^{\pi\kappa}\!=\!0
\label{4}\\
\nonumber
&\!\!\frac{1}{2}[\frac{1}{4}(\Omega\!-\!R)^{\alpha\nu\rho}\varepsilon_{\alpha\nu\rho\mu}\varepsilon^{\pi\kappa\mu\sigma}+\\
&+g^{\sigma\kappa}(\Omega\!-\!R)^{\pi a}_{\phantom{\mu a}a}]M_{\pi\kappa}\!-\!mU^{\sigma}\!=\!0
\label{5}
\end{eqnarray}
in terms of the vector current and the tensor. If the above Weyl equations were to be taken with no electrodynamics nor torsion they would be completely identical to the first three Majorana equations in the case of no mass, and in both cases we witness that the field equations are reduced to a form that results to be genuinely algebraic.

Some comment for the singular spinors of the last three classes of Lounesto classification: the fourth class is the most general and we know no solution; in the Weyl case there are exact solutions in presence of gravitation as it is described in \cite{Cianci:2015pba}; the Majorana case is somewhat special because spinors of this type in recent times have been the subject of a number of studies \cite{daRocha:2013qhu, daSilva:2012wp, Ablamowicz:2014rpa, daRocha:2016bil, daRocha:2008we, Villalobos:2015xca, daRocha:2011yr, Cavalcanti:2014uta, daRocha:2007sd}, and in particular the theory of eigen-spinors of charge conjugation has been developed in terms of spinors called ELKO (German for \textit{Eigenspinoren des LadungsKonjugationsOperators}), in a self-contained way \cite{Ahluwalia:2004sz,Ahluwalia:2004ab}. We will talk about ELKO next.
\section{Operation of Charge Conjugation, ELKO and Neutrino Masses}
The Majorana spinor is defined to be an eigen-spinor of the operation of charge conjugation $\eta\boldsymbol{\gamma}^{2}\psi^{\ast}\!=\!\psi$ with $\eta$ an irrelevant constant phase, and because this expression in German has the acronym ELKO, this is the name that has been recently given to the theory that presupposes to study them in the most detailed manner: there is still a difference between the standard Majorana spinors and these non-standard ELKO spinors, and that is that Majorana spinors are Grassmann-valued fields undergoing to a first-order differential field equation while ELKO spinors are defined in terms of a new dual and they undergo to a second-order differential field equation. In the Majorana case their being Grassmann-valued has the same role that in the ELKO case the new dual has, namely they are the elements needed to have the field equations derived from a Lagrangian. But it is intrinsic to their structure that a Majorana and an ELKO entail different Lagrangians.

More in detail, for the Majorana spinors, the general form of field is such that $i\boldsymbol{\gamma}^{2}\psi^{\ast}\!=\!\psi$ or explicitly as
\begin{eqnarray}
&\!\!\psi\!=\!\left(\!\begin{tabular}{c}
$i\boldsymbol{\sigma}^{2}\phi^{\ast}$\\
$\phi$
\end{tabular}\!\right)\!=\!\left(\!\begin{tabular}{c}
$Be^{-i\beta}$\\
$-Ae^{-i\alpha}$\\
$Ae^{i\alpha}$\\
$Be^{i\beta}$
\end{tabular}\!\right)
\end{eqnarray}
where $\phi$ is a semi-spinor of defined chirality while $A$, $B$, $\alpha$, $\beta$ are four real scalar fields; this spinor might be seen as a potential solution of the Dirac field equation
\begin{eqnarray}
&i\boldsymbol{\gamma}^{\mu}\boldsymbol{\nabla}_{\mu}\psi\!-\!m\psi\!=\!0
\end{eqnarray}
which for $q\!=\!0$ is invariant under charge conjugation as expected; this should come from the Lagrangian
\begin{eqnarray}
&\mathscr{L}\!=\!\frac{i}{2}(\overline{\psi}\boldsymbol{\gamma}^{\mu}\boldsymbol{\nabla}_{\mu}\psi
\!-\!\boldsymbol{\nabla}_{\mu}\overline{\psi}\boldsymbol{\gamma}^{\mu}\psi)
\!-\!m\overline{\psi}\psi
\end{eqnarray}
if it were not for the fact that in this case $\overline{\psi}\psi\!\equiv\!0$ identically, a problem that might be solved by requiring that they be Grassmann-valued fields. For ELKO spinors, the construction is similar from the kinematical point of view in the sense that they are such that $\boldsymbol{\gamma}^{2}\lambda^{\ast}\!=\!\pm\lambda$ or
\begin{eqnarray}
&\lambda\!=\!\left(\!\begin{tabular}{c}
$\pm\boldsymbol{\sigma}^{2}\phi^{\ast}$\\
$\phi$
\end{tabular}\!\right)\!=\!\left(\!\begin{tabular}{c}
$\mp iBe^{-i\beta}$\\
$\pm iAe^{-i\alpha}$\\
$Ae^{i\alpha}$\\
$Be^{i\beta}$
\end{tabular}\!\right)
\end{eqnarray}
differing from a Majorana spinor only up to an irrelevant constant phase and where the double sign corresponds to self-conjugate and antiself-conjugate ELKO; in the most general approach one can define a new dual for ELKO as
\begin{eqnarray}
&\widetilde{\lambda}\!=\!i\lambda^{\dagger}\boldsymbol{\gamma}^{0}\varepsilon
\end{eqnarray}
where $\varepsilon$ is the matrix that flips the helicity of the spinor as described by equation (4.55) of \cite{Ahluwalia:2016rwl}; a common result widely reviewed in ELKO literature is that ELKO fields thus constructed are not solutions of the Dirac equation, so that ELKO fields are postulated to be potential solutions of the Klein-Gordon--like field equation
\begin{eqnarray}
&\boldsymbol{\nabla}^{2}\lambda\!+\!m^{2}\lambda\!=\!0
\end{eqnarray}
invariant under charge conjugation and coming from
\begin{eqnarray}
&\mathscr{L}\!=\!\boldsymbol{\nabla}^{\alpha}\widetilde{\lambda}\boldsymbol{\nabla}_{\alpha}\lambda
\!-\!m^{2}\widetilde{\lambda}\lambda
\end{eqnarray}
in which $\widetilde{\lambda}\lambda\!\neq\!0$ in general \cite{Ahluwalia:2016rwl}. Consequently, the problem that for the Majorana spinors has been circumvented by introducing Grassmann variables for ELKO spinors is not even there because of Ahluwalia duals; and Majorana fields satisfy a first-order derivative field equation while ELKO fields do not and thus second-order derivative field equations seem needed. For a good comparison between the two situations see also the references \cite{daRocha:2007pz,HoffdaSilva:2009is}.

We discuss now how we think Majorana and ELKO to be more similar than what is usually believed. For sake of clarity, we will use the above formalism very heavily.

We shall start from the fact that, as we have discussed above, it is always possible to find a frame in which some given Majorana spinor field can be written as
\begin{eqnarray}
&\!\!\psi\!=\!\left(\!\begin{tabular}{c}
$1$\\
$0$\\
$0$\\
$1$
\end{tabular}\!\right)
\label{Majorana}
\end{eqnarray}
and thus with no real scalar degree of freedom; in return, this means that Majorana spinor fields saturate the entire Lounesto class, and because ELKO are in the same class, then ELKO must be Majorana spinor fields, and not only kinematically, but in general dynamics: since the above results have been obtained by exploiting transformation properties of the bi-linear fields, which depend on the Dirac dual, and because for an ELKO it is the Ahluwalia dual that has to be considered, one might argue that the previous derivation would no longer be valid, and therefore we will demonstrate our claim by direct inspection, employing transformation laws. We recall that the transformation laws for spinorial fields are given by
\begin{eqnarray}
\boldsymbol{S}_{R1}\!=\!\left(\begin{array}{cccc}
\!\cos{\theta}\!&\!i\sin{\theta}\!&\!0\!&\!0\!\\ 
\!i\sin{\theta}\!&\!\cos{\theta}\!&\!0\!&\!0\!\\ 
\!0\!&\!0\!&\!\cos{\theta}\!&\!i\sin{\theta}\!\\ 
\!0\!&\!0\!&\!i\sin{\theta}\!&\!\cos{\theta}\!\\
\end{array}\right)
\end{eqnarray}
for the rotation around the first axis of angle $\theta$ with
\begin{eqnarray}
\boldsymbol{S}_{R2}\!=\!\left(\begin{array}{cccc}
\!\cos{\varphi}\!&\!\sin{\varphi}\!&\!0\!&\!0\!\\ 
\!-\sin{\varphi}\!&\!\cos{\varphi}\!&\!0\!&\!0\!\\ 
\!0\!&\!0\!&\!\cos{\varphi}\!&\!\sin{\varphi}\!\\ 
\!0\!&\!0\!&\!-\sin{\varphi}\!&\!\cos{\varphi}\!\\
\end{array}\right)
\end{eqnarray}
for the rotation around the second axis of angle $\varphi$ and
\begin{eqnarray}
\boldsymbol{S}_{R3}\!=\!\left(\begin{array}{cccc}
\!e^{i\zeta}\!&\!0\!&\!0\!&\!0\!\\ 
\!0\!&\!e^{-i\zeta}\!&\!0\!&\!0\!\\ 
\!0\!&\!0\!&\!e^{i\zeta}\!&\!0\!\\ 
\!0\!&\!0\!&\!0\!&\!e^{-i\zeta}\!\\
\end{array}\right)
\end{eqnarray}
for the rotation around the third axis of angle $\zeta$ while
\begin{eqnarray}
&\boldsymbol{S}_{B3}\!=\!\left(\begin{array}{cccc}
\!e^{-\eta}\!&\!0\!&\!0\!&\!0\!\\ 
\!0\!&\!e^{\eta}\!&\!0\!&\!0\!\\ 
\!0\!&\!0\!&\!e^{\eta}\!&\!0\!\\ 
\!0\!&\!0\!&\!0\!&\!e^{-\eta}\!\\
\end{array}\right)
\end{eqnarray}
is the boost along the third axis of rapidity $\eta$ in the most general case; the general form of ELKO fields is given by
\begin{eqnarray}
&\lambda\!=\!\left(\!\begin{tabular}{c}
$-iBe^{-i\beta}$\\
$iAe^{-i\alpha}$\\
$Ae^{i\alpha}$\\
$Be^{i\beta}$
\end{tabular}\!\right)
\end{eqnarray}
where we have specialized for the self-conjugate (of course the antiself-conjugate would give the same result): if we calculate $\lambda'\!=\!\boldsymbol{S}_{R1}\boldsymbol{S}_{R2}\lambda$ we see that for $\lambda'$ we may choose to simultaneously set the second and third components to zero (or we may choose to simultaneously set the first and fourth components to zero) if and only if
\begin{eqnarray}
\nonumber
&Ae^{i\alpha}(\cos{\theta}\cos{\varphi}\!-\!i\sin{\varphi}\sin{\theta})+\\
&+Be^{i\beta}(\cos{\theta}\sin{\varphi}\!+\!i\cos{\varphi}\sin{\theta})\!=\!0
\end{eqnarray}
and since this relationship can aways be inverted in order to get in terms of $A/B$ and $\alpha\!-\!\beta$ the angles $\theta$ and $\varphi$ then we conclude that it is always possible to write ELKO as
\begin{eqnarray}
&\lambda'\!=\!\left(\!\begin{tabular}{c}
$e^{-i(\beta+\frac{\pi}{2})}$\\
$0$\\
$0$\\
$e^{i\beta}$
\end{tabular}\!\right)\!B
\end{eqnarray}
in general; if now we calculate $\lambda''\!=\!\boldsymbol{S}_{R3}\lambda'$ we see that it is always possible to pick in terms of $\beta$ the function $\zeta$ so that we could always write ELKO like
\begin{eqnarray}
&\lambda''\!=\!\left(\!\begin{tabular}{c}
$1$\\
$0$\\
$0$\\
$1$
\end{tabular}\!\right)\!B
\end{eqnarray}
up to $e^{i\frac{\pi}{4}}$ which is an irrelevant constant phase; finally, if we calculate $\lambda'''\!=\!\boldsymbol{S}_{B3}\lambda''$ we see that it is always possible to pick in terms of $B$ the function $\eta$ so that we can always write ELKO according to the expression
\begin{eqnarray}
&\lambda'''\!=\!\left(\!\begin{tabular}{c}
$1$\\
$0$\\
$0$\\
$1$
\end{tabular}\!\right)
\end{eqnarray}
which recovers a result by Ahluwalia for which in the case of ELKO the boost does not mix the various components but it simply amounts to an overall scalar shift \cite{Ahluwalia:2016rwl}, and eventually proving that ELKO can always be reduced to a Majorana field in the form (\ref{Majorana}) in general. This proves mathematically that ELKO and Majorana spinor fields are the same kinematically, and because this proof does not suppose spinor fields to be constant, but it is done in general in the case in which they are point-dependent fields, then there is nothing that prevents these results to be extended to their full dynamical behaviour.

What this means in particular is that there should in principle be nothing preventing ELKO to be solutions of first-order derivative field equations of Dirac type; on the other hand, it is an established result that ELKO are not solutions of the Dirac equation \cite{Ahluwalia:2016rwl}: our interpretation of this apparent conundrum is that albeit Ahluwalia's initial calculations are obviously correct, nevertheless they are valid only for ELKO in form of plane waves as in \cite{Ahluwalia:2016rwl} but not in general. In fact, if we take ELKO as considered in terms of plane waves, they could be reduced to
\begin{eqnarray}
&\!\!\psi\!=\!\boldsymbol{S}\!\left(\!\begin{tabular}{c}
$1$\\
$0$\\
$0$\\
$1$
\end{tabular}\!\right)
\end{eqnarray}
just in terms of rigid $\boldsymbol{S}$ so that $R_{ij\mu}\!=\!0$ and in addition there is also no gravitational field: hence the Majorana spinor field equations (\ref{1}, \ref{2}, \ref{3}, \ref{4}, \ref{5}) reduce to the mere constraint $m\!=\!0$ compatibly with results reviewed in reference \cite{Ahluwalia:2016rwl}. So we confirm that ELKO fields in the form of plane waves cannot be solution of flat space-time Dirac equations, but we do not stretch this up to stronger claims for which any ELKO however general cannot be a solution of correspondingly general Dirac equations.

It is fundamental to remark here that ELKO spinors as potential solutions for some types of Dirac spinor field equations is not a new idea, and some work can be found in literature such as for instance in reference \cite{VazJr:2016jnp}.

We stress however that this does not mean that ELKO theories should be rejected, and in fact we think that the contrary is true: ELKO theories, with the definition of the Ahluwalia dual, do provide the missing ingredient for the Majorana theory to be completed. The problem with a Majorana field is that its dynamical field equations can be derived from a Lagrangian only if we can compute a non-vanishing bi-linear scalar: this is a thing that does not seem possible, unless we employ some supplementary structure like the Grassmann variables, but Grassmann variables, despite being a well-defined concept, appear to be included as a patch over a gap of a pre-existing theory, a solution that might look artificial to some. Without the help of some super-imposed additional structure, it would seem impossible to give the definition of a non-vanishing bi-linear scalar, or better, we do not have a non-vanishing bi-linear scalar if the conjugated spinor is the usual Dirac conjugated spinor. But if we make use of the Ahluwalia conjugated spinor then a non-vanishing bi-linear scalar is clearly possible as we have discussed just above.

The spinor conjugation is not uniquely defined and as a matter of fact there are at least two that can be found, namely the standard Dirac conjugation and the recently introduced Ahluwalia conjugation. As Dirac conjugation, with its flipping of chiral components, is best suited for eigen-spinors of parity, Ahluwalia conjugation, with its flipping of helicity as well as chiral components, is best suited for eigen-spinors of charge conjugation. Once each type of spinor has the spinorial conjugation that suites it most, the construction of the Lagrangian is as usual.

There is now a problem which, with all these elements, seems to be less tragic, when seen in a new light. 

The problem is that of the neutrino mass generation mechanism. It is in fact a well established experimental result that neutrinos display oscillation, the feature for which neutrino flavours are transmuted from one another across the neutrino propagation: if neutrinos had masses, neutrino oscillations could be explained. Contrary to the common belief, there is no known proof that the reverse implication also works, and so we have to keep an open mind for the possibility that some mechanism can fit the oscillations even in absence of mass, but still, we do know that neutrino masses do imply neutrino oscillation, and so it might be simpler to have a mechanism of neutrino mass generation. This mechanism, however, is elusive.

In the standard model of particle physics, neutrinos are taken to be massless, and although with the inclusion of right-handed sterile neutrinos it is possible to generate a neutrino mass along with the lepton mass, in exactly the same manner in which we generate both quark masses, this solution suffers for the fact that right-handed sterile neutrinos do not seem to exist; the lack of right-handed neutrinos is not a problem if we can create them starting from left-handed neutrinos, and Majorana fields do this precisely, although now the problem is that the Majorana mass is $\mathscr{L}_{\mathrm{Majorana}}\!=\!Y\phi^{\dagger}L\overline{L}\phi$ (with $L$ being the doublet of left-handed spinors and $\phi$ being the doublet of complex scalars), which is a term of mass dimension $5$ that spoils the property of renormalizability of the standard model in its core. So far, there does not seem to be a way out, but a possible solution may reside in the Ahluwalia dual used for the neutrinos. The Lagrangian of the standard model in the material sector would then be given by
\begin{eqnarray}
\nonumber
&\mathscr{L}_{\mathrm{matter}}\!=\!
i\overline{L}\boldsymbol{\gamma}^{\mu}\boldsymbol{\nabla}_{\mu}L
\!+\!i\overline{R}\boldsymbol{\gamma}^{\mu}\boldsymbol{\nabla}_{\mu}R
\!+\!i\widetilde{\lambda}\boldsymbol{\gamma}^{\mu}\boldsymbol{\nabla}_{\mu}\lambda+\\
\nonumber
&+\nabla_{\mu}\phi^{\dagger}\nabla^{\mu}\phi
\!+\!\Lambda^{2}(v^{2}\phi^{2}\!-\!\frac{1}{2}\phi^{4})-\\
&-Y_{e}(\overline{L}\phi R\!+\!\overline{R}\phi^{\dagger}L)
\!-\!Y_{\nu}(\overline{L}i\boldsymbol{\sigma}^{2}\phi^{\ast}\lambda
\!-\!\widetilde{\lambda}\phi^{T}i\boldsymbol{\sigma}^{2}L)
\label{L}
\end{eqnarray}
where the transformations of the fields are the standard gauge transformations of the $SU(2)_{L}\!\times\!U(1)$ group
\begin{eqnarray}
&\lambda'=\lambda\\
&R'=e^{-i\alpha}R\\
&L'=e^{-\frac{i}{2}\left(\vec{\boldsymbol{\sigma}} \cdot \vec{\theta}
+\alpha\mathbb{I}\right)}L\\
&\phi'=e^{-\frac{i}{2}\left(\vec{\boldsymbol{\sigma}} \cdot \vec{\theta}-\mathbb{I}\alpha\right)}\phi
\end{eqnarray}
so that Lagrangian (\ref{L}) would be fully gauge invariant.

After the spontaneous symmetry breaking and having gone into the unitary gauge, where $\phi^{T}\!=\!(0,v)$ is given as the Higgs vacuum, we can write the Yukawa potential as
\begin{eqnarray}
\nonumber
&\!\!\!\!\!\!\!\!\mathscr{L}^{\mathrm{Yukawa}}_{\mathrm{matter}}\!=\!
-Y_{e}v[(e_{L})^{\dagger}e_{R}\!+\!(e_{R})^{\dagger}e_{L}]-\\
&-Y_{\nu}v[i(\varphi_{L\downarrow})^{T}\!\boldsymbol{\sigma}^{2}\nu_{L\uparrow}]
\label{Ls}
\end{eqnarray}
having taken
\begin{eqnarray}
&\!\!\lambda\!=\!\left(\!\begin{tabular}{c}
$\varphi_{L\uparrow}$\\
$\boldsymbol{\sigma}^{2}(\varphi_{L\uparrow})^{\ast}$
\end{tabular}\!\right)
\end{eqnarray}
and
\begin{eqnarray}
&\!\!R\!=\!\left(\!\begin{tabular}{c}
$0$\\
$e_{R}$
\end{tabular}\!\right)
\end{eqnarray}
with
\begin{eqnarray}
&\!\!L\!=\!\left(\!\begin{tabular}{c}
$\left(\!\begin{tabular}{c}
$\nu_{L\uparrow}$\\
$0$
\end{tabular}\!\right)$\\
$\left(\!\begin{tabular}{c}
$e_{L}$\\
$0$
\end{tabular}\!\right)$
\end{tabular}\!\right)
\end{eqnarray}
in which the up arrow $\uparrow$ stands for the spin-up helicity, and exhibiting both a Dirac mass term and an Ahluwalia mass term. Explicitly, it is possible to see that
\begin{eqnarray}
&\!\!\widetilde{\lambda}\!=\!i\left(\begin{array}{c|c}
(\varphi_{L\downarrow})^{T}\boldsymbol{\sigma}^{2}\ &\ (\varphi_{L\downarrow})^{\dagger}
\end{array}\right)
\end{eqnarray}
in which the down arrow $\downarrow$ is for the spin-down helicity, and in which we have that $i(\varphi_{L\downarrow})^{T}\!\boldsymbol{\sigma}^{2}\nu_{L\uparrow}\!\neq\!0$ in the most general of the possible circumstances as it can be seen by direct inspection. Lagrangian (\ref{Ls}) has two non-null mass terms, both come from a Yukawa potential, whose mass dimension $4$ ensures renormalizability, none involves any right-handed sterile neutrino, never observed so far.

In the struggle between restrictions of renormalizability and the need to construct a mass term, the dual first introduced by Ahluwalia releases the tension by offering a viable neutrino mass generation mechanism.
\section{Conclusion}
In this paper, we have reviewed the classification of the spinors between regular and singular, and among the singular we have better studied the separation between Weyl and Majorana spinors; in this last case, we have detailed the discussion on eigen-spinors of the charge conjugation operation, highlighting analogies and differences between standard Majorana spinors and new ELKO spinors: the main difference is that the former are Grassmann-valued while the latter have to satisfy a higher-order derivative field equation. We have hence proceeded to show how it may be thinkable for ELKO to satisfy least-order derivative field equations of Dirac type and that we do not need to develop a theory of Majorana spinors with Grassmann variables if we define the Ahluwalia dual: the Ahluwalia dual is another form of dual that can be defined alongside to the Dirac dual, the two differing for the fact that the Ahluwalia dual can be employed to build bi-linear tensors mixing helicity and chirality states while the Dirac dual can be used to build bi-linear tensors that mix chirality states only. As we recognize that not one but two duals can be introduced, each best suited for a specific type of spinors, a solution for the problem of the neutrino mass generation mechanism is seen to arise quite naturally.

In fact, the neutrino mass generation mechanism has the problem that the neutrino mass term is either of Dirac type, and so it would need right-handed sterile neutrinos still undetected, or of Majorana type, and hence it is non-renormalizable: this appears to be unsolvable, and it is unsolvable if we stick to the use of Dirac dual. But if we pass to the employment of the Ahluwalia dual, we have a simple way of considering only the left-handed standard neutrinos to build some renormalizable mass term.

Would this Ahluwalia mass term still give rise to the same type of neutrino oscillations? And if not, what are the main differences that this alternative mass term, with its mixing of helicity states, could bring about?

A mechanism of chiral oscillations in connection with flavor oscillations has also been studied \cite{Bernardini:2007ew,Bernardini:2005wh}. And an alternative mechanism of mass generation has also been presented recently in the literature \cite{Bernardini:2012sc,Bernardini:2007ex}.

These works contain some parallel study with respect to our investigations, so how would our results compare to those of Bernardini and colleagues?

\end{document}